\documentclass[11pt]{article}\usepackage[]{graphicx}\usepackage[]{color}
\makeatletter
\def\maxwidth{ %
  \ifdim\Gin@nat@width>\linewidth
    \linewidth
  \else
    \Gin@nat@width
  \fi
}
\makeatother

\definecolor{fgcolor}{rgb}{0.345, 0.345, 0.345}

\usepackage{framed}
\makeatletter
 {\par\unskip\endMakeFramed%
 \at@end@of@kframe}
\makeatother

\definecolor{shadecolor}{rgb}{.97, .97, .97}
\definecolor{messagecolor}{rgb}{0, 0, 0}
\definecolor{warningcolor}{rgb}{1, 0, 1}
\definecolor{errorcolor}{rgb}{1, 0, 0}

\usepackage{alltt}
\usepackage[utf8]{inputenc}
\usepackage{amsmath}
\usepackage{amsfonts}
\usepackage{amscd}
\usepackage{amssymb}
\usepackage{natbib}
\usepackage{url}
\usepackage{graphicx,times}
\usepackage{setspace}
\usepackage{ragged2e}

\usepackage{geometry}
\IfFileExists{upquote.sty}{\usepackage{upquote}}{}
\begin{document}

\begin{center}

\noindent {\huge \bf Challenging nostalgia and performance metrics in baseball} \\

\vspace{.2in}

Daniel J. Eck

\vspace{.01in}

\textit{Department of Biostatistics, Yale University, New Haven, CT, 06510. \\ \url{daniel.eck@yale.edu}}

\vspace{.1in}

\end{center}



\section{Introduction}
\doublespacing
It is easy to be blown away by the accomplishments of great old time 
baseball players when you look at their raw or advanced baseball statistics.  
These players produced mind-boggling numbers. For example, see 
Babe Ruth's batting average and pitching numbers, 
Ty Cobb's 1911 season, 
Walter Johnson's 1913 season, 
Tris Speaker's 1916 season, 
Rogers Hornsby's 1925 season, 
and
Lou Gehrig's 1931 season.
The statistical feats achieved by these players (and others) far surpass 
the statistics that recent and current players produce.  At first glance 
it seems that players from the old eras were vastly superior to the 
players in more modern eras, but is this true? 
In this paper, we investigate whether baseball players from earlier 
eras of professional baseball are overrepresented among the game's all-time 
greatest players according to popular opinion, performance metrics, and expert 
opinion.  We define baseball players from earlier eras to be those that 
started their MLB careers in the 1950 season or before.  
This year is chosen because it coincides with the decennial US Census 
and is close to 1947, the year in which baseball became integrated. 

In this paper we do not compare baseball players via their statistical 
accomplishments.  Such measures exhibit era biases that are confounded with 
actual performance.  Consider the single season homerun record as an example. 
Before Babe Ruth, the single season homerun record was 27 by Ned Williams in 
1884. 
Babe Ruth broke this record in 1919 
when he hit 29 homeruns.  He subsequently destroyed his own record in 
the following 1920 season when he hit 54 homeruns.  The runner up in 1920 
finished the season with a grand total of 15 homeruns.  At this point in time 
homerun hitting was not an integral part of a batter's approach. 
This has changed. Now, we often see multiple batters reach at least 30-40 
homeruns within one season and a 50 homerun season is not a rare 
occurrence. 
In the 1920s, Babe Ruth stood head and shoulders above his peers due to a 
combination of his innate talent and circumstance.  
His approach was quickly emulated and became widely adopted. 
However, Ruth's accomplishments as a homerun hitter would not stand out nearly 
as much if he played today and put up similar homerun totals.    
The example of homeruns hit by Babe Ruth and the impact they had relative 
to his peers represents a case where adjustment towards a peer-derived 
baseline fails across eras.  No one reasonably expects 1920 Babe Ruth to hit 
more than three times the amount of homeruns hit by the second best homerun 
hitter if the 1920 version of Babe Ruth played today.  
This is far from an isolated case.  

There are several statistical approaches currently used to compare baseball 
players across eras. 
These include 
wins above replacement as calculated by baseball reference (bWAR), 
wins above replacement as calculated by fangraphs (fWAR), 
adjusted OPS+, 
adjusted ERA+, 
era-adjusted detrending \citep{petersen}, 
computing normal scores as in Jim Albert's work on a Baseball Statistics Course 
in the Journal of Statistics Education, 
and era bridging \citep{berry1999eras}. 
A number of these are touted to be season adjusted and the remainder are 
widely understood to have the same effect.  
In one way or another all of these statistical approaches compare the  
accomplishments of players within one season to a baseline that 
is computed from statistical data within that same season.  
This method of player comparison ignores talent discrepancies that exist across 
seasons as noted by Stephen J. Gould in numerous lectures and papers.
Currently, there is no definitive quantitative or qualitative basis for 
comparing these baselines, which are used to form intra-season player 
comparisons, across seasons.  These methods therefore fail to properly 
compare players across eras of baseball despite the claim that they are 
season adjusted.  

Worse still is that these approaches exhibit a favorable bias towards baseball 
players who played in earlier seasons 
\citep{schmidt2005concentration}.  
We explore this bias from two separate theoretical perspectives underlying how 
baseball players from different eras would actually compete against each 
other.  The first perspective is that players would teleport across eras to 
compete against each other.  From this perspective, the players from earlier eras 
are at a competitive disadvantage because, on average, baseball players have 
gotten better as time has progressed.
Specifically, it is widely acknowledged that 
fastball velocity, pitch repertoire, training methods, and management 
strategies have all improved over time.  
We do not find the teleportation perspective to be of 
much interest for these reasons.  The second perspective is that a player from 
one era could adapt naturally to the game conditions of another era if they 
grew up in that time. 
This line of thinking is challenging to current statistical methodology because 
adjustment to a peer-derived baseline no longer makes sense. 
Even in light of these challenges with the second perspective, we find that the 
players from earlier eras are overrepresented among baseball's all time greats.  
We justify our findings through the consideration of population dynamics which 
have changed drastically over time.  


\section{Data}

The MLB eligible population is not well-defined.  As a proxy, we can say 
that the MLB eligible population is the decennial count of males aged 
20-29 that are living in the United States (US) 
and Canada. 
Baseball was segregated on racial grounds until 1947.  As a result, 
African American and Hispanic American population counts in the US  
and Canada are added to our dataset starting in 1960.  The year 1960 is chosen 
because the integration of the MLB was slow as noted in Armour's work on the 
integration of baseball in the Society for American Baseball Research.


Players from Central and South American countries and the Caribbean islands 
were also targets of discrimination.
We have added data from these countries to the MLB eligible population starting 
in 1960:
Mexico, 
the Dominican Republic, 
Venezuela, 
Cuba, 
Panama, 
Puerto Rico, 
Netherlands Antilles, 
Aruba, 
Honduras, 
Jamaica, 
the Bahamas, 
Peru, 
Columbia, 
Nicaragua, 
and the United States Virgin Islands.  
In the mid to late 1990s, the MLB and minors saw an influx of Asian baseball 
players from Japan, South Korea, Taiwan, and the Philippines.  
We have added the populations of these countries to the MLB eligible population 
starting in 2000.  
In 2010, the MLB established a national training center in Brazil 
as noted in Lor{\'e}'s work on the popularity of baseball in Brazil in the 
Culture Trip.
Therefore, we have included the Brazilian population of 20-24 year old 
men  
into our MLB eligible population starting in 2010.  
We estimate that the 2011-2015 MLB eligible population is half of the 
MLB eligible population counted in the 2010 decennial Censuses.  We expect 
that this underestimates the actual 2011-2015 MLB eligible population 
since we have observed a constant increase in the overall MLB eligible 
population as time increases.

The MLB eligible population is displayed in Table 1.
The cumulative proportion means that at each era, the population of the 
previous eras is also included. As an example of how to interpret this 
dataset, consider the year 1950. There were 11.59 
million males aged 20-29. The proportion of the historical MLB eligible 
population that existed at or before 1950 is 0.187.



\begin{table}[ht]
\centering
\begin{tabular}{lccc}
  \hline
 & year & population & cumulative population proportion \\ 
  \hline
1 & 1880 & 4.440 & 0.013 \\ 
  2 & 1890 & 5.010 & 0.027 \\ 
  3 & 1900 & 5.580 & 0.043 \\ 
  4 & 1910 & 8.560 & 0.068 \\ 
  5 & 1920 & 8.930 & 0.093 \\ 
  6 & 1930 & 9.920 & 0.122 \\ 
  7 & 1940 & 11.130 & 0.154 \\ 
  8 & 1950 & 11.590 & 0.187 \\ 
  9 & 1960 & 18.420 & 0.240 \\ 
  10 & 1970 & 24.490 & 0.310 \\ 
  11 & 1980 & 33.930 & 0.407 \\ 
  12 & 1990 & 37.460 & 0.515 \\ 
  13 & 2000 & 60.660 & 0.689 \\ 
  14 & 2010 & 72.270 & 0.896 \\ 
  15 & 2015 & 36.140 & 1.000 \\ 
   \hline
\end{tabular}
\caption{Eligible MLB population throughout the years. The first column 
    indicates the year, the second column indicates the estimated MLB eligible 
    population size (in millions), and the third column indicates the proportion 
    of the MLB eligible population in row x that was eligbile at or before row x.} 
\end{table}

\section{The greats}

To determine which players are the all-time greatest players, we consult four 
lists which reflect popular opinion, performance metrics, and expert opinion 
that purport to determine the greatest players.  The first 
list is compiled by Ranker, 
which is constructed entirely from popular opinion as determined by up and 
down votes.  
The second and third lists rank players by highest career WAR as calculated 
by baseball reference and fangraphs, respectively. 
The fourth list is a ranking from ESPN 
and is based on expert opinion and statistics. 

The rankings for all four lists are given in Table~\ref{top25}.  
As an example of the information contained in Table~\ref{top25} consider 
the greatest players of all time according to ESPN  
displayed in the fourth column.  
We see that 5 players that started their careers before 1950 are in the top 10 
all time and 11 players that started their careers before 1950 are in the top 
25 all time.  When the MLB eligible population is considered, it appears that 
the players from the earlier eras are overrepresented in this particular list.  

\begin{table}[h!]
\begin{center}
\begin{tabular}{lllll}
\hline
rank & Ranker & bWAR & fWAR & ESPN \\
\hline
1  & {\bf Babe Ruth}         & {\bf Babe Ruth}      & {\bf Babe Ruth}      & {\bf Babe Ruth}      \\
2  & {\bf Ty Cobb}           & {\bf Cy Young}       & Barry Bonds          & Willie Mays          \\
3  & {\bf Lou Gehrig}        & {\bf Walter Johnson} & Willie Mays          & Barry Bonds          \\
4  & {\bf Ted Williams}      & Barry Bonds          & {\bf Ty Cobb}        & {\bf Ted Williams}   \\
5  & {\bf Stan Musial}       & Willie Mays          & {\bf Honus Wagner}   & Hank Aaron           \\
6  & Willie Mays             & {\bf Ty Cobb}        & Hank Aaron           & {\bf Ty Cobb}        \\
7  & Hank Aaron              & Hank Aaron           & Roger Clemens        & Roger Clemens        \\
8  & Mickey Mantle           & Roger Clemens        & {\bf Cy Young}       & {\bf Stan Musial}    \\
9  & {\bf Rogers Hornsby}    & {\bf Tris Speaker}   & {\bf Tris Speaker}   & Mickey Mantle        \\
10 & {\bf Honus Wagner}      & {\bf Honus Wagner}   & {\bf Ted Williams}   & {\bf Honus Wagner}   \\
11 & {\bf Cy Young}          & {\bf Stan Musial}    & {\bf Rogers Hornsby} & {\bf Lou Gehrig}     \\
12 & {\bf Walter Johnson}    & {\bf Rogers Hornsby} & {\bf Stan Musial}    & {\bf Walter Johnson} \\
13 & {\bf Joe Dimaggio}      & {\bf Eddie Collins}  & {\bf Eddie Collins}  & Greg Maddux          \\
14 & Sandy Koufax            & {\bf Ted Williams}   & {\bf Walter Johsnon} & Rickey Henderson     \\ 
15 & Ken Griffey Jr.         & {\bf Pete Alexander} & Greg Maddux          & {\bf Rogers Hornsby} \\
16 & {\bf Jimmie Foxx}       & Alex Rodriguez       & {\bf Lou Gehrig}     & Mike Schmidt         \\
17 & {\bf Tris Speaker}      & {\bf Kid Nichols}    & Alex Rodriguez       & {\bf Cy Young}       \\
18 & {\bf Joe Jackson}       & {\bf Lou Gehrig}     & Mickey Mantle        & Joe Morgan           \\
19 & Mike Schmidt            & Rickey Henderson     & Randy Johnson        & {\bf Joe Dimaggio}   \\
20 & Nolan Ryan              & Mickey Mantle        & {\bf Mel Ott}        & Frank Robinson       \\
21 & {\bf Christy Mathewson} & Tom Seaver           & Nolan Ryan           & Randy Johnson        \\
22 & Roberto Clemente        & {\bf Mel Ott}        & Mike Schmidt         & Tom Seaver           \\
23 & Albert Pujols           & {\bf Nap Lajoie}     & Rickey Henderson     & Alex Rodriguez       \\
24 & {\bf Cap Anson}         & Frank Robinson       & Frank Robinson       & {\bf Tris Speaker}   \\
25 & Greg Maddux             & Mike Schmidt         & Burt Blyleven        & Steve Carlton        \\
 & & & & \\
pre-1950 in top 10 &   7 \,/\, 10  &   6 \,/\, 10  &   6 \,/\, 10  &   5 \,/\, 10  \\
pre-1950 in top 25 &  15 \,/\, 25  &  15 \,/\, 25  &  12 \,/\, 25  &  11 \,/\, 25  \\
\hline
\end{tabular}
\end{center}
\caption{Lists of the top 25 greatest baseball players to ever play in the 
  MLB according to Ranker.com (1st column), bWAR (2nd column), 
  fWAR (3rd column), and ESPN (4th column). Players that started their career 
  before 1950 are indicated in bold. The last two rows count the number of players 
  that started their careers before 1950 in each of the top 10 and top 25 lists 
  respectively.}
\label{top25}
\end{table}

\section{Statistical evidence}
\label{sec:Stats}

We now provide evidence that the top 10 and top 25 lists displayed in 
Table~\ref{top25} overrepresent players who started their careers 
before 1950.  
We require two assumptions for the validity of our calculations which we will 
explore in detail in the next Section. 
These assumptions are: 
\begin{itemize}
\item First, we assume that innate talent is uniformly distributed over the 
  MLB eligible population over the different eras.
\item Second, we assume that the outside competition to the MLB available by 
  other sports leagues after 1950 is offset by the increased salary 
  incentives received by MLB players.
\end{itemize}

With these assumptions in mind we calculate the probability that at least x 
people from each top 10 and top 25 list in Table~\ref{top25} started their 
career before 1950 using the proportion depicted in Table 1.  Consider the 
bWAR list for example.  According to bWAR, we see that 6 of the top 10 
players started their careers before 1950.  From Table 1 we see that the 
proportion of the MLB eligible population that played at or  
before 1950 was approximately 0.187.  
We then calculate the probability that one would expect to observe 6 or more 
individuals in a top 10 list from that time period where the chance of 
observing each individual is about 0.187.  We calculate 
this probability using the Binomial distribution.  
We perform the same type of extreme event 
calculation for each top 10 and top 25 list depicted in Table~\ref{top25}.  
The results are provided in Table~\ref{probvalues}.

\begin{table}[h!]
\begin{center}
\begin{tabular}{lllll}
\hline
  &  Ranker  &  bWAR  &  fWAR  &  ESPN \\
  \hline
probability of extreme event in top 10 list 
  & 0.000562 
  & 0.00448 
  & 0.00448 
  & 0.0249 \\
probability of extreme event in top 25 list 
  & 0.0000057 
  & 0.0000057 
  & 0.000826 
  & 0.00322 \\
chance of extreme event in top 10 list 
  & 1 in 1780 
  & 1 in 223 
  & 1 in 223 
  & 1 in 40 \\
chance of extreme event in top 25 list 
  & 1 in 174816 
  & 1 in 174816 
  & 1 in 1210 
  & 1 in 310 \\
  \hline
\end{tabular}
\end{center}
\caption{The probability and chance (1 in 1/probability) of each extreme event 
  calculation corresponding to the four lists in Table~\ref{top25}.}
\label{probvalues}
\end{table}

As an example of how to interpret the results of Table~\ref{probvalues}, 
continue with bWAR's top 10 list.  Table~\ref{probvalues} shows that the 
probability of observing 6 or more players that started their careers at 
or before 1950 of the top 10 all time players, based on population 
dynamics, is about 0.00448 
(a chance of 1 in 223).
The same interpretation applies to remainder of Table~\ref{probvalues}.  
The results provided in Table~\ref{probvalues} present overwhelming evidence 
that players who started their careers before 1950 are overrepresented in top 
10 and top 25 lists from the perspectives of fans, analytic assessment of 
performance, and experts' rankings.  

\section{Assumptions and Sensitivity Analysis}
\label{sec:Assumptions}

The results in Table~\ref{probvalues} are valid 
under the two assumptions provided in the previous Section.  In the first of 
these assumptions we specify that innate talent is evenly dispersed across 
eras. 
We do not fully believe that the first assumption holds because the 
distribution of innate talent has improved over time as the MLB eligible  
population has expanded as noted by Stephen J. Gould,  
Christina Kahrl at ESPN, and in 
Martin B. Schmidt and David J. Berri's work on concentration of baseball 
talent in the Journal of Sports Economics.
This suggests that the probabilities displayed in Table~\ref{probvalues} are 
conservative.  

With respect to the second assumption, we note that the 
National Basketball Association (NBA) and the National Football League (NFL) 
started in 1946 and 1920 respectively 
with both sports greatly rising in popularity since the inception of their 
respective professional leagues.  Soccer and hockey have also risen in 
popularity in the United States.  That being said, it is widely known that 
MLB salaries have substantially increased.
For example, the 1967 census lists the median US household income as \$7,200. 
The minimum MLB salary at that time was \$6,000 as noted by the LA Times 
sports writer Bill Shaiken in a piece titled ``A look at how Major League 
Baseball salaries have grown by more than 20,000\% the last 50 years.''
In short,
baseball players made far less than they do today relative to the general US 
population and it is unlikely that one could consider playing professional 
baseball to be a lucrative career in the earlier eras. 
These figures offer evidence that while other 
professional leagues may have drawn from the MLB eligible talent pool, 
salary incentives have led to an increase in the overall quality of MLB 
players.  

Though we cannot confirm this theory with absolute certainty, at worst, our 
our second assumption suffers some modest violations.  
To account for this possibility we consider a sensitivity 
analysis applied to the findings in Table~\ref{probvalues}.  We weight the 
decennial populations displayed in Table 1 to reflect the overall interest 
that the US population has had in baseball over time irrespective of salary 
increases based on Gallup polling data.  
The four weighting regimes that we consider 
are given in Table 4 below.  
These regimes serve as proxies for the proportion of the 
MLB eligible population thought to strive towards a career in professional 
baseball.  
In an effort to be conservative, we have deliberately placed greater weight 
on the time periods before 1940 for each weighting regime because no polling 
data is available.
We do not expect the MLB eligible population before 1940 to be as 
high as our weighting regimes suggest because of 
relatively modest baseball attendance figures in early eras of baseball,  
non existence of the radio prior to 1920, 
the dead-ball era,
and low compensation.  



David W. Moore and Joseph Carroll's Gallup article entitled 
``Baseball Fan Numbers Steady, But Decline May Be Pending'' shows that 
interest in baseball has remained steady since 1937, at approximately
40\%.  
Consistent with this benchmark, the first and second weighting regimes 
(w1 and w2) conservatively place 0.50 and 0.60 weights, repectively, 
on fan interest prior to 1940.  
The third weighting regime (w3), constructed from the Gallup polling data 
(\url{https://news.gallup.com/poll/4735/sports.aspx}), 
reflects the proportion of the US population who 
listed baseball as their favorite sport.  
The appropriateness of this regime is intuitively questionable because 
some people play baseball even if it is not their favorite sport and the 
weight placed on pre-1940 years is very high.  
The fourth weighting regime (w4) is the average of w2 and w3.  


These weights are obtained from survey data from the US because similar data 
is unavailable from other countries.
We applied these same 
weights to all of the other countries, even though interest in baseball in 
these other countries is thought to either be on par with or much greater 
than the US.  Therefore our weighting regimes address, and in fact, 
overcompensate for any potential shortcomings of no weighting.

\begin{table}[ht]
\centering
\begingroup\footnotesize
\begin{tabular}{rrrrrrrrrrrrrrrr}
  \hline
 & 1880 & 1890 & 1900 & 1910 & 1920 & 1930 & 1940 & 1950 & 1960 & 1970 & 1980 & 1990 & 2000 & 2010 & 2015 \\ 
  \hline
w1 & 0.50 & 0.50 & 0.50 & 0.50 & 0.50 & 0.50 & 0.40 & 0.40 & 0.40 & 0.40 & 0.40 & 0.40 & 0.40 & 0.40 & 0.40 \\ 
  w2 & 0.60 & 0.60 & 0.60 & 0.60 & 0.60 & 0.60 & 0.40 & 0.40 & 0.40 & 0.40 & 0.40 & 0.40 & 0.40 & 0.40 & 0.40 \\ 
  w3 & 0.40 & 0.40 & 0.40 & 0.40 & 0.40 & 0.40 & 0.35 & 0.38 & 0.34 & 0.28 & 0.16 & 0.16 & 0.13 & 0.12 & 0.10 \\ 
  w4 & 0.50 & 0.50 & 0.50 & 0.50 & 0.50 & 0.50 & 0.38 & 0.39 & 0.37 & 0.34 & 0.28 & 0.28 & 0.27 & 0.26 & 0.25 \\ 
   \hline
\end{tabular}
\endgroup
\caption{Weighting regimes.} 
\end{table}

\begin{table}[h!]
\begin{center}
\begin{tabular}{llllll}
\hline
 weight & &  Ranker  &  bWAR  &  fWAR  &  ESPN \\
 \hline
w1 & probability of extreme event in top 10 list 
  & 0.00121 
  & 0.00839 
  & 0.00839 
  & 0.0406 \\
& probability of extreme event in top 25 list 
  & 0.0000267 
  & 0.0000267 
  & 0.0025 
  & 0.00845 \\
& chance of extreme event in top 10 list 
  & 1 in 824 
  & 1 in 119 
  & 1 in 119 
  & 1 in 25 \\
& chance of extreme event in top 25 list 
  & 1 in 37519 
  & 1 in 37519 
  & 1 in 401 
  & 1 in 118 \\
  & & & & & \\
w2 & probability of extreme event in top 10 list 
  & 0.0023 
  & 0.0141 
  & 0.0141 
  & 0.0604 \\
& probability of extreme event in top 25 list 
  & 0.0000944 
  & 0.0000944 
  & 0.00608 
  & 0.0182 \\
& chance of extreme event in top 10 list 
  & 1 in 434 
  & 1 in 71 
  & 1 in 71 
  & 1 in 17 \\
& chance of extreme event in top 25 list 
  & 1 in 10595 
  & 1 in 10595 
  & 1 in 164 
  & 1 in 55 \\
  & & & & & \\
w3 & probability of extreme event in top 10 list 
  & 0.0311 
  & 0.109 
  & 0.109 
  & 0.273 \\
& probability of extreme event in top 25 list 
  & 0.0128 
  & 0.0128 
  & 0.152 
  & 0.266 \\
& chance of extreme event in top 10 list 
  & 1 in 32 
  & 1 in 9 
  & 1 in 9 
  & 1 in 3.7 \\
& chance of extreme event in top 25 list 
  & 1 in 78 
  & 1 in 78 
  & 1 in 6.6 
  & 1 in 3.8 \\
  & & & & & \\  
w4 & probability of extreme event in top 10 list 
  & 0.00622 
  & 0.0311 
  & 0.0311 
  & 0.11 \\
& probability of extreme event in top 25 list 
  & 0.000649 
  & 0.000649 
  & 0.0227 
  & 0.0561 \\
& chance of extreme event in top 10 list 
  & 1 in 161 
  & 1 in 32 
  & 1 in 32 
  & 1 in 9.1 \\
& chance of extreme event in top 25 list 
  & 1 in 1542 
  & 1 in 1542 
  & 1 in 44 
  & 1 in 18 \\
  \hline
\end{tabular}
\end{center}
\caption{The probability and chance (1 in 1/probability, rounded) 
  of each extreme event calculation corresponding to the four lists in 
  Table~\ref{top25} after the MLB eligible population in Table 1 is 
  weighted according to the four conservative weighting regimes.}
\label{probvalues.weights}
\end{table}

Table~\ref{probvalues.weights} shows the effect of these weighting regimes as 
applied to the results in Table~\ref{probvalues}.  
The conclusions from weighting populations with respect to w1, w2, and w4 in 
Table~\ref{probvalues.weights} are largely consistent with those in 
Table~\ref{probvalues}.  
The third weighting regime presents some conflicting conclusions.  When 
weighting populations with respect to w3 we see that popular opinion and 
bWAR overrepresent players who started their careers before 1950.  
However, the same is not so for fWAR and ESPN. 
The overall finding of this sensitivity analysis is that conservatively 
weighting populations with respect to fan interest in baseball yields the 
conclusion as the analysis in Section 4:
it is very unlikely that the pre-1950s time period could 
have produced so many historically great baseball players.

\section{Additional comparison methods}


\subsection{Versus your peers methods}
\label{WARcritique}

There are several methods which are used to compare players across eras that 
do so by computing a baseline achievement threshold within one season and then 
comparing players to that baseline.  These methods then rank players by how far 
they stood above their peers, the greatest players were better than their peers 
by the largest amount. 
We have shown that this approach can exhibit major biases in player comparisons 
as evidenced by career bWAR and fWAR.  Adjusted OPS+ is a worse offender 
than bWAR or fWAR.  Adjusted ERA+ is right in line with ESPN rankings.

\subsection{PPS detrending}

We describe and critique the methodology of \citet{petersen} (PPS). 
As described in PPS, they detrend player statistics by normalizing 
achievements to seasonal averages, which they claim accounts for changes in 
relative player ability resulting from both exogenous and endogenous factors, 
such as 
talent dilution from expansion, 
equipment and training improvements, 
as well as performance enhancing drug usage. 
PPS misunderstands the effect of talent dilution from expansion and ignores 
reality.  The talent pool was more diluted in the earlier eras of 
baseball than now because of a small relative eligible population size and 
the exclusion of entire populations of people on racial grounds.  
See Table~\ref{dilution} for the specifics.  PPS's position with respect 
to equipment and training improvements is likewise not without fault 
because the same improvements are equally available to every competitor.  
Finally, PPS does not account for increases in salary compensation enjoyed by 
MLB players in modern eras, and their methodology fails to address 
segregation prior to 1947.

\begin{table}[h!]
\begin{center}
\begin{tabular}{lcccc}
\hline
year & eligible pop. & number of teams & roster size & eligible pop. per roster spot \\
\hline
1890 & 5.01  & 8  & 15 & 41.7   \\
1910 & 8.56  & 16 & 25 & 21.4  \\
1930 & 9.92  & 16 & 25 & 24.8  \\
1950 & 11.59  & 16 & 25 & 29  \\
1970 & 24.49  & 24 & 25 & 40.8  \\
1990 & 37.46 & 26 & 25 & 57.6 \\
2010 & 72.27 & 30 & 25 & 96.4 \\
\hline
\end{tabular}
\end{center}
\caption{Relative talent dilution when considering the MLB eligible population 
  sizes at select time periods. Eligible population totals are in millions in 
  column 2 and are in thousands in column 5. }
\label{dilution}
\end{table}

The mathematics of PPS detrending is also questionable in the context of 
comparing baseball players across eras. 
PPS notes that the evolutionary nature of competition results in a 
non-stationary rate of success.  They then detrend player 
statistics by normalizing achievements to seasonal averages.  
The normalization goes as follows: 
	Suppose a player hits 40 homeruns in a given season and that the league 
	average prowess for homerun hitting in that season is 10 homeruns. If the 
	historical average prowess for homerun hitting is 5 homeruns then our 
	player's detrended homerun count in that particular season is 
	$40\times(5/10) = 20$.  In general, the detrending formula is 
	$Y \times (\text{historic prowess} / \text{league prowess})$ where $Y$ is 
  individual prowess for a particular player in a given season.
We see PPS detrending as an inflationary metric of relative prowesses 
and not a detrending metric.  
Fundamentally different approaches for detrending are advocated in 
authoritative textbooks such as 
 Introduction to Time Series and Forecasting,
   by
 Peter J. Brockwell and Richard A. Davis.
Table 2 in PPS displays the top 25 career detrended homerun totals.  
It is clear that having higher prowess relative to your peers, hitting more 
runs in this case, is not indicative of a player's prowess with respect to 
peers from fundamentally different eras.  

\subsection{Era bridging}

\citet{berry1999eras} claim that their era bridging technique accounts for 
talent discrepancies across eras.  However, they do not explicitly 
parameterize this in their hierarchical models.  They state that 
``globalization has been less pronounced in the MLB (relative to other 
sports)... Baseball has remained fairly stable within the 
United States, where it has been an important part of the culture for more 
than a century'' \citep{berry1999eras}.  This rationale ignores 
segregation, increases in the MLB eligible population 
relative to available roster spots, and increases in the average overall 
talent of that population.  
Therefore, there methodology does not fully address the characteristics 
of a changing talent pool.  

In \citet[panel (c) of Figure 7]{berry1999eras} we see that their model 
predicts that a .300 hitter in 1996 will have a lower than .300 average for 
several seasons from 1900-1920.  This conflicts with the well-established 
notion that the talent of baseball players has improved over time.  
In \citet[Table 9]{berry1999eras} we see that 6 of the 10 best hitters 
for average started their career before 1950 and 10 of the 25 best hitters 
for batting average started their careers before 1950.  Their paper was 
published in 1999 so we recompute the chances of these events where the MLB 
eligible population ends at 1999. 
We calculate the chance that one would expect to 
observe 6 or more individuals in a top 10 list who started their careers 
before 1950 as 1 in 30. 
We calculate the chance that one would expect to observe 10 or more 
individuals in a top 25 list who started their careers before 1950 as 
1 in 7.7.  These chances are not as extreme 
as those in Table~\ref{probvalues}, but they still correspond to events 
that are unlikely.

\section{Conclusions}



The MLB players from the early eras of baseball receive significant attention 
and praise as a result of their statistical achievements and their mythical 
lore.  We find that these players are collectively overrepresented in 
rankings of the greatest players in the history of the MLB, and that popular 
performance metrics such as WAR fail to properly compare players across eras.  
Superior statistical accomplishments achieved 
by players that started their careers before 1950 are a reflection of our 
inability to properly compare talent across eras.  It is highly unlikely that 
athletes from such a scarcely populated era of available baseball talent 
could represent top 10 and top 25 lists so abundantly. 

We close with a general discussion on greatness.  The conclusions of this 
article have broader implications than just rankings of baseball players.  
Who are the greatest all-time athletes in other sports, 
artists, 
musicians, 
actors and actresses,
scientists, or 
leaders?  Do our perceptions change when we focus beyond nostalgia?  

\section*{Acknowledgements}
This work would not be what it is today without the many conversations that 
the author had with family, friends, and peers.
I am very grateful to 
Jim Albert, 
Peter M. Aronow, 
James Burrell, 
Xiaoxuan Cai, 
R. Dennis Cook, 
Forrest W. Crawford, 
Steven A. Culpepper, 
Andrew Depuy, 
Evan Eck, 
Fred Eck, 
Kim Eck, 
Marcus A. Eck, 
Michael Eck, 
Phil Eck, 
Shirley Eck, 
Wes Eck, 
Margret Erlendsdottir, 
Soheil Eshghi, 
Charles J. Geyer, 
Ed Kaplan, 
Zehong (Richard) Li, 
Adam Maidman, 
Aaron Molstad, 
Olga Morozova, 
Oliver Om, 
Kerry Purcell, 
Ken Ressel, 
Erick Ruuttila, 
Jesse Ruuttila, 
Anne Schuh, 
Bill Schuh, 
Yushuf Sharker, 
Ben Sherwood, 
Stephanna Szotkowski,
Dootika Vats,
Brandon Whited,
the editorial board at Chance, 
and 
two anonymous referees at Chance
for helpful comments and discussions (some of which went very long).
This work was partially supported by NIH grants NICHD DP2 HD091799-01.

\begin{FlushLeft}

\end{FlushLeft}

\end{document}